# Vision as an Energy-Driven Process

Robert Pepperell[1]

Cardiff Metropolitan University,

200 Western Avenue,

Cardiff, CF5 2YB, UK

## Abstract

It is frequently assumed that the fundamental purpose of vision is to extract information from light and process it in the brain in order to gain knowledge about environmental objects. Treating vision as an information-driven process has been fruitful for many areas of vision science. But this approach has inherent limitations and may not be sufficient to explain at the most fundamental level why and how we see. Energy is a prime mover in the evolution and operation of all biological systems, including vision, and early researchers in physiological optics and experimental psychology regarded vision as an energy-driven process. Yet to date there has been little attempt to analyze the fundamentals of vision in energetic terms. This paper is a provisional attempt to sketch an account of the early stages of vision as an energy-driven process. On this account, vision is a form of biological work in which energy from the environment is absorbed and processed by neurobiological systems in order to guide behaviour. Evidence from the evolution and neurobiology of vision is presented in support of this approach. I conclude that treating vision as an energy-driven process can contribute to our understanding of its fundamental nature and operation.

## 1. Introduction

What is the fundamental purpose of vision? How is light from the world transformed into visual experiences by physical processes in eyes and brains? Scientific approaches to such foundational questions often assume that vision is driven by the acquisition and processing of information (Diamant, 2008). David Marr's influential paper entitled 'Visual Information Processing' (Marr, 1980) talks of the "different kinds of information in a visual scene." A standard textbook on vision defines it as "…the process of acquiring knowledge about environmental objects and events by extracting information from the light they emit or reflect" (Palmer, 1999). The "information rate" of data transmission between the human eye and the brain is calculated to be roughly that of an ethernet connection (Koch et al., 2006). Photoreceptive cells, such as rods, are said to function as processors of "visual information" that "transmit 1 or 0" (Sterling & Laughlin, 2015).

[1] Corresponding author: rpepperell@cardiffmet.ac.uk

Treating vision as an information-driven process has proved fruitful for many fields of inquiry, particularly where the powerful mathematical tools derived from information theory have been employed to model the inherently complex processes involved (Delgado-Bonal & Martín-Torres, 2016; Shannon, 1948). But as will be discussed, the notion of information as currently used in science has certain limitations when applied to the operation of vision at a fundamental level. We therefore need to explore other approaches if we are to explain what vision does and how it works.

Energy is widely acknowledged to be a foundational property of reality (Smil, 2017) and the dynamical flow of energy drives all living processes (Chaisson, 2002; Dyson, 1971; Gibson, 2013; Morowitz, 1979; Eric D. Schneider & Sagan, 2005; Smil, 2017; Stoffregen & Bardy, 2001; Wong-Riley, 2010). Some of the most significant early researchers in neurobiology, perceptual psychology, and physiological optics regarded perception and cognition as being fundamentally energy-driven (Adelmann & Sherrington, 1942; Fechner, 1966; Helmholtz, 1924).[2] One might therefore expect that energy's role in the neurobiology of vision—including visual perception and cognition—would be widely and keenly studied. But to date there has been little attempt to analyse the fundamental nature of vision in dynamic terms. This paper is a provisional attempt to sketch an account of vision as an energy-driven process.[3]

---

[2] Helmholtz opens the third volume of his *Treatise on Physiological Optics* with the statement: "Perceptions of external objects being therefore of the nature of ideas, and ideas themselves being invariably activities of our psychic energy, perceptions also can only be the result of psychic energy" (p. 1). Fechner devotes chapter V of *Elements of Psychophysics* to discussing the nature of energy and its foundational role in physics and perception. He says: "Just as it takes a certain quantity of kinetic energy to split a log or lift a given weight to a given height, so does it take a certain quantity to think a thought of a given intensity; and energy for one can be changed into energy for the other" (p. 36).

[3] Kuzma (Kuzma, 2019) has recently presented an energy-focused account of visual perception, while Wong-Riley (Wong-Riley, 2010) has addressed the metabolism of the early visual system. Though nominally a theory about the role of free energy in sensation, perception and action, Karl Friston's work is primarily information-theoretical and provides an account of sensory behaviour based on principles of machine learning

Two main lines of evidence are discussed here: the role of energy in the evolution of vision and the energetic nature of the physical and neurobiological processes involved in early vision. Although discussion will be limited to a few elementary facts of physics and neurophysiology, the purpose is to highlight some of the principles that distinguish this approach from the information-driven one, and to outline how they can contribute to our understanding of the fundamental nature and purpose and operation of vision.

## 2. The Nature of Energy

Energy is conventionally defined in physics as the capacity to do work and work is the act of applying force to move objects against an opposing force (Law & Rennie, 2015; Smith, 1998). Energy can take many forms, but all are either kinetic or potential. In both cases, energy derives its motive power—its capacity to do work—from the *differences* between energy states within a system. For kinetic energy it is the difference in motion between one part of a system and another, as in the case of a bat having a certain mass and velocity that strikes a ball having lesser mass and velocity, so transferring some energy and causing the ball to move faster. For potential energy it is the difference in position, or the tension, between one part of the system and another, as in the case of a ball that is held above the ground, where the greater the difference in height the greater the potential energy. Physical systems have a natural tendency to equalize or minimize these energy state differences—sometimes called gradients or potentials—and the forces released during the act of minimization can be harnessed to do work. An example of this 'difference principle' is the harnessing of minimizations of temperature differences within a steam engine to drive a piston (see Fig. 1 for this and other examples).

Forces and work can also be understood in fundamental terms as manifestations of difference: forces consist in antagonistic pairs of pushes or pulls, like those existing between positively and negatively charged subatomic particles, and work brings about a difference in the velocity and position of an object, as when a ball is lifted from the ground. Thinking of energy, forces, and work as manifestations of difference, I suggest, can help us to appreciate

---

and Bayesian modelling. 'Free energy' is defined in this information-theoretic context as an amount of surprisal in a system, or a variational bound on informational entropy (Friston, 2008).

their intrinsic nature and how they operate integrally and dynamically. This will be important when we come to analyze the critical role that differences play in the physics and biology of vision. But the differences manifested by energy, forces and work are not merely conceptual; they are physical properties that can be individually and precisely quantified using SI units such as the joule, newton, or volt and the change in the free energy in a chemical reaction can be determined using the Helmholtz or Gibbs formulae.[4] Moreover, as energy differences interact in complex ways in physical systems they can give rise to dynamic behaviours and hierarchically organized structures that we can measure and model using mathematics and the statistical tools of information theory—as will be discussed.[5]

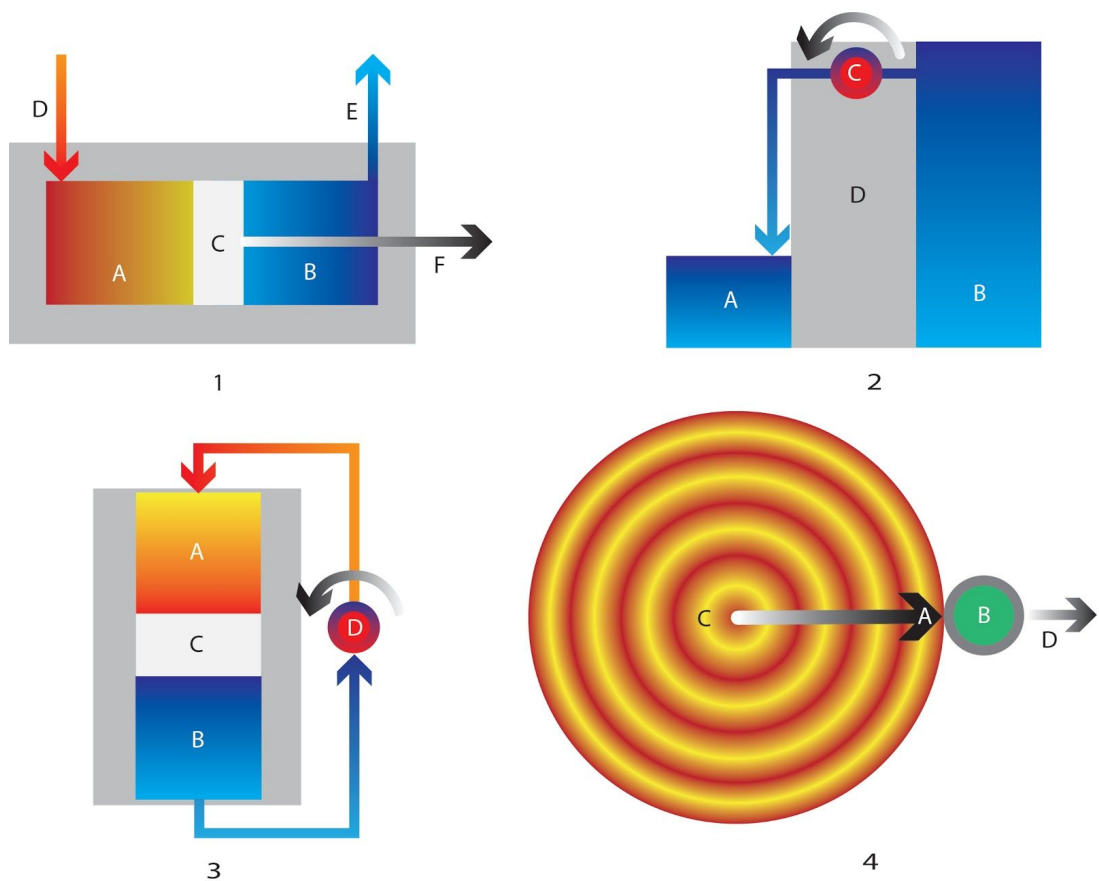

**Fig. 1.** Some examples of the 'difference principle' according to which energy can drive work. 1. *A (non-cyclic) steam engine*: The engine consists of a chamber (A) containing vapour that is at a higher

---

temperature (greater kinetic energy) than that in (B), having been heated by a source (D); the vapour in (A) expands and exerts a greater force on a piston (C) than in (B) and as this inequality tends towards equilibrium so the difference of temperature within the system is minimized. The piston is forced to move, expelling the cooler vapour through (E), and providing a source of work (F). 2. *A turbine dam*: The water in the reservoir (A) is lower than in (B), which is retained by the dam (D); the water at (B) tends towards equilibrium with (A) and when provided with a conduit flows to minimize the difference; in doing so it drives a turbine (C) that can power work. 3. *An electric battery*: The battery contains a positively charged cathode at (A) and a negatively charged anode (B) separated by a barrier at (C); once connected via a suitable conductor, the difference in charge between (A) and (B) will tend to minimize, causing electrons to flow, the motion of which is harnessed to drive a motor (D) and so do work. 4. *Light (electromagnetic) energy*: A source of illumination (C) radiates waves with a given velocity; when these waves at (A) come into contact with an object having a lower velocity (B) they transfer some of their energy, so minimizing the difference between the velocities and doing work on the object (D). In all these cases it is the *difference* between the energy state of the system at (A) and at (B)—sometimes referred to as the gradient or potential—that provides the motive power to drive the work. (Colour figures required)

## 3. Energy and the Evolution of Vision

Vision emerged during the Cambrian period some 540 million years ago and its arrival is thought to have profoundly determined the path of all subsequent evolution (Parker, 2004). Like most living processes, the evolution of vision depended on energy generated by the nuclear reaction occurring in the sun some 93 million miles away and carried to earth by electromagnetic waves (Land & Fernald, 1992; Smil, 2017). The visible part of this solar energy occupies a narrow band of the electromagnetic spectrum at wavelengths of between approximately 400 to 700 nm, and comprises some 40% of the energy arriving from the sun. This energy can interact with and alter the molecular structure of certain chemicals on earth, triggering reactions that drive organic processes such as photosynthesis, photoreception, and phototransduction. These processes depend not just on the arrival of light energy but on the organic matter on earth being in a lower energy state than the light it interacts with. For, as discussed above, it is the difference between these energy states—between, for example, the molecular compounds in a chloroplast and the photons it absorbs—that provides the motive power to drive this organic work.[6]

---

[6] The nightly cooling of the part of the earth's surface facing away from the sun, when lower grade energy (long-wave radiation) is emitted into space, is critical to the maintenance of life as it replenishes the difference

Precambrian organisms lacked eyes but were still able to exploit light energy for survival. Photoreceptive compounds helped early cells to avoid the harmful effects of high-energy ultraviolet radiation, to repair DNA, and capture and store energy from the sun (Schwab et al., 2012). Early unicellular organisms drifted with the flow of their substrates, and the gradual acquisition of locomotory capability was a major evolutionary step as it increased the chances of finding favourable conditions or avoiding harmful ones (Jékely, 2009). But powering the mechanisms for detecting environmental conditions and locomotion also required an energy supply. The rhodopsin family of compounds proved especially successful for this need, and their incorporation into cellular chemistry gave early life forms a significant competitive advantage by enabling them to transduce the energy carried by light.

Some forms of rhodopsin can act as a biological battery by pumping protons across a cell membrane to create an electrochemical gradient that stores energy. This potential energy can be converted into kinetic energy through structures such as the flagellum that propel and steer the cell through its environment. Euglena gracilis is a common example of a single-cell organism that exploits the energy in sunlight and the transducing properties of rhodopsin to drive propulsion (Schwab et al., 2012). The origins of vision, therefore, are intimately linked to driving behaviour.

As eyes evolved, the electrochemical energy transduced by rhodopsin was used to initiate ever more complex neurological processing which, along with the development of brains, supported increasingly sophisticated behaviours or tasks. According to recent researchers (Land & Nilsson, 2012) (Nilsson, 2013) the evolution of vision proceeded through four major steps: non-directional sensitivity, directional sensitivity, low spatial resolution, and high spatial resolution, each supporting a different class of visually guided task. Each of these steps, which will now be described, is also a way in which life learnt to harness not only the motive power of the energy differences—the gradients or potentials described above—but also the spatiotemporal differences of wavelength, intensity and direction of propagation that in energy states that enable life to harness the high grade light energy (short-wave radiation) during the day. This dependence on a cyclical process of energy absorption and radiation is what Harold Morowitz refers to as 'energy flow': "...it is not energy *per se* that makes life go, but the flow of energy through the system" (Morowitz, 1979).

occur *within* the light energy—the 'differences of differences'—as will now be exemplified in each of the four evolutionary steps:

I.  *Non-Directional sensitivity.* Early organisms were able to respond to the presence or absence of ambient light, or the intensity of light, using unscreened photoreceptors, but could not determine its direction. Nevertheless, in addition to the uses noted above, non-directional light responses allowed organisms to control circadian rhythms and photoperiodicity, burrowing behaviour, detect predators by cast shadows, avoid harmful levels of UV radiation, and detect water depth (Land & Nilsson, 2012). The absence or presence of light is a simple example of a difference of differences, that is, a change in the global amount of energetic differences available to an organism.

II. *Directional sensitivity.* Innovations that allowed creatures to determine the direction of light relative to their own bodies included photoreceptor screening pigments and body movements for scanning the environment. These expanded the range of light-dependent behaviours, including greater light energy harvesting, more degrees of locomotory freedom, control of body posture and orientation, and alarm responses for detecting predators (Jékely, 2009). This further example of a difference of differences is due to changes in the direction of propagation of the energetic differences as detected by the organism.

III. *Low spatial resolution.* True vision emerged with the advent of light focusing lenses able to project an image onto arrays of photoreceptors (Schwab et al., 2012).[7] Although equipped with relatively poor optics and few photoreceptors, the resolving power and direction sensing properties of the first true eyes supported a greatly increased repertoire of behaviours, including self-detection of motion, object avoidance, habitat selection, and spatial orientation in relation to environmental or celestial landmarks. In this case, the differences in the differences being detected were the varying patterns or structures contained in the array of photons detected by the organism, the nature of which depended on the source of illumination and its interactions with objects in the environment.

---

[7] The focus here is on camera-type eyes, but similar low-resolution visual capability is afforded by compound eyes.

IV.  *High spatial resolution.* The emergence of high-resolution spatial vision, with larger mosaics of photoreceptors and improved optics, again expanded the range of behaviours that could be guided by sight, including the pursuit of prey and fight from predators at larger distances, mate detection and attraction, fine spatial orientation, visual communication, and social interaction. Humans benefit from high resolution vision, which has contributed greatly to their evolutionary success. In this case, the visual system is better able to differentiate the differences of quantity, direction, colour and structure in the global array of photons, that is, to maximally differentiate the differences of differences.

Fig. 2 illustrates a hierarchical arrangement of these differences of differences, organized according to the four evolutionary steps. Note that each layer in the diagram consists of differences that are also at the same time *integrated* insofar as they constitute parts of a single system or process within that layer. Differences of differences are effectively *integrated differences* insofar as they are integrated into a single system or structure. These integrated differences form ever more complex structures and patterns that display features like regularity, similarity, and variation.

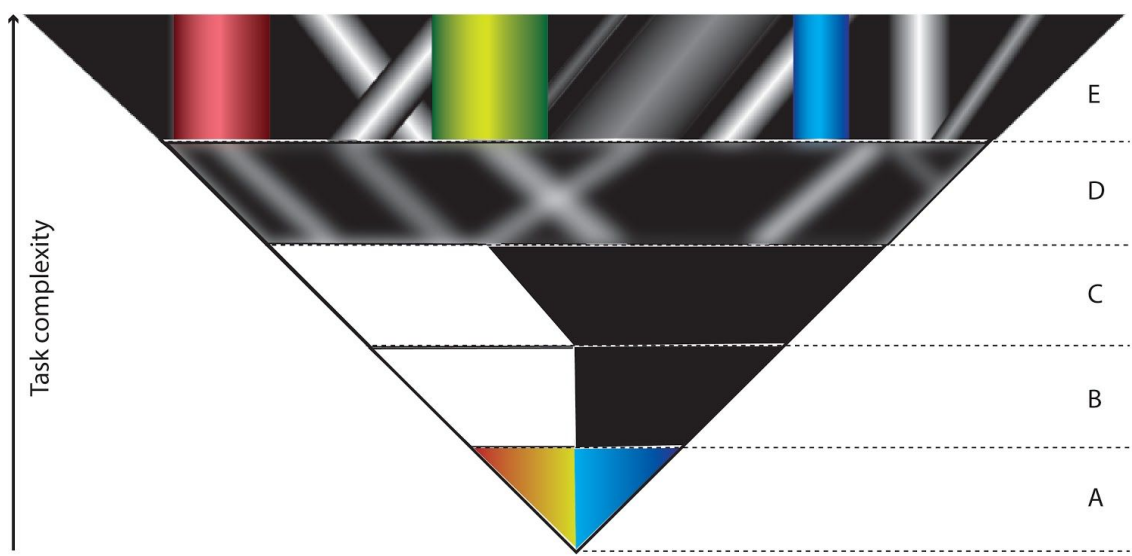

**Fig. 2.** An illustration of the hierarchy of differences (differences of differences, or integrated differences) pertinent to the evolution of vision. Layer A represents the fundamental energy difference (gradient or potential) that provides the motive power to do work on the sensory systems of the organism; the higher energy state is depicted in orange and the lower state in blue. Layer B represents what can be sensed with the most rudimentary photosensitive apparatus—the difference between presence and absence of light. This is a primary level of integration, where the energy differences in A, which are able to do work on the organism, are differentiated by their presence or absence. Layer C represents the difference of light direction relative to an organism, indicated by the slanted boundary, which can be sensed with more sophisticated photosensitive apparatus. Layer D represents what more complex photodection systems such as true eyes can sense, that is, multiple differences in the structure of the light that are caused by interactions between the source of illumination and objects in the environment, as indicated by several slanted regions at different angles and intensities, with the blurriness indicating low resolution discrimination. Layer E represents what more advanced eyes can detect, which is a complex structure composed of differences of colour, direction, and intensity, the sharpness representing high resolution discrimination. As vision evolved by exploiting the motive power of the energy difference represented by A it acquired ever greater capacity to differentiate the differences contained in the structure of the light, from B in the simplest life-forms with a minimal repertoire of tasks to E in more complex ones, such as humans, with a vast repertoire of behaviours. Here is illustrated only the spatial dimension of difference; equally important is the temporal dimension—the way light differs over time.

Michael Land and Dan-Eric Nilsson (Land & Nilsson, 2012) stress that the evolution of vision is driven by the need to support ever more complex visually guided tasks. Executing these tasks depends on responding advantageously to variations in the electromagnetic energy received by the organism from its environment; this is the primary work of all vision. Following this, we can identify three key evolutionary trends that gradually led to the highly complex visual systems and behavioural repertoires that humans, and other animals with similar visual systems, enjoy:

1) *Increasing the quantity of energy received.* Increasing the quantity of photons captured by the eye increases the statistical reliability of the image processing that differentiates external objects (Pirenne, 1967). Enhanced photon capture is achieved by the growth of absolute size of the eye, the increase in the number of photoreceptors, the stacking of photoreceptors in layers that trap more passing photons, faster photoreceptive chemistry, and movement strategies for eyes and bodies (Land & Nilsson, 2012; Nilsson, 2013).

2) *Increasing the acuity of the differentiation of the energy.* Differentiating between the incoming arrays of photons with greater acuity enhances detection of objects in the environment and aids finer motor responses. Greater acuity is achieved by increasing the overall number of photoreceptors and the surface area of the photoreceptive layer onto which a larger image can be projected. Human eyes developed a macular region populated with densely packed receptors and a sophisticated musculature to align the macula with regions of interest in space. Acuity is also enhanced by improving the optics of the eye so that a sharply focused image is projected onto the photoreceptive layer, and onto the macula in particular.

3) *Increasing integration of the differences in energy.* Increasing the acuity of local energy differentiation at the photoreceptive layer necessitates an increase in the capacity for integrating those differences into more globally organized patterns within the animal's neurobiology. This entails an expansion of the visual processing capabilities of the brain and greater coordination with sensory and motor functions (Nilsson, 2009). Comparing patterns of energetic activity across the photoreceptive layer using bipolar and ganglion cells, for example, enables the visual system to extract features such as contours, contrasts and movement by neurologically summing differentiated values and passing them on for further processing.[8]

---

[8] Such patterns would conventionally be referred to as 'information'.

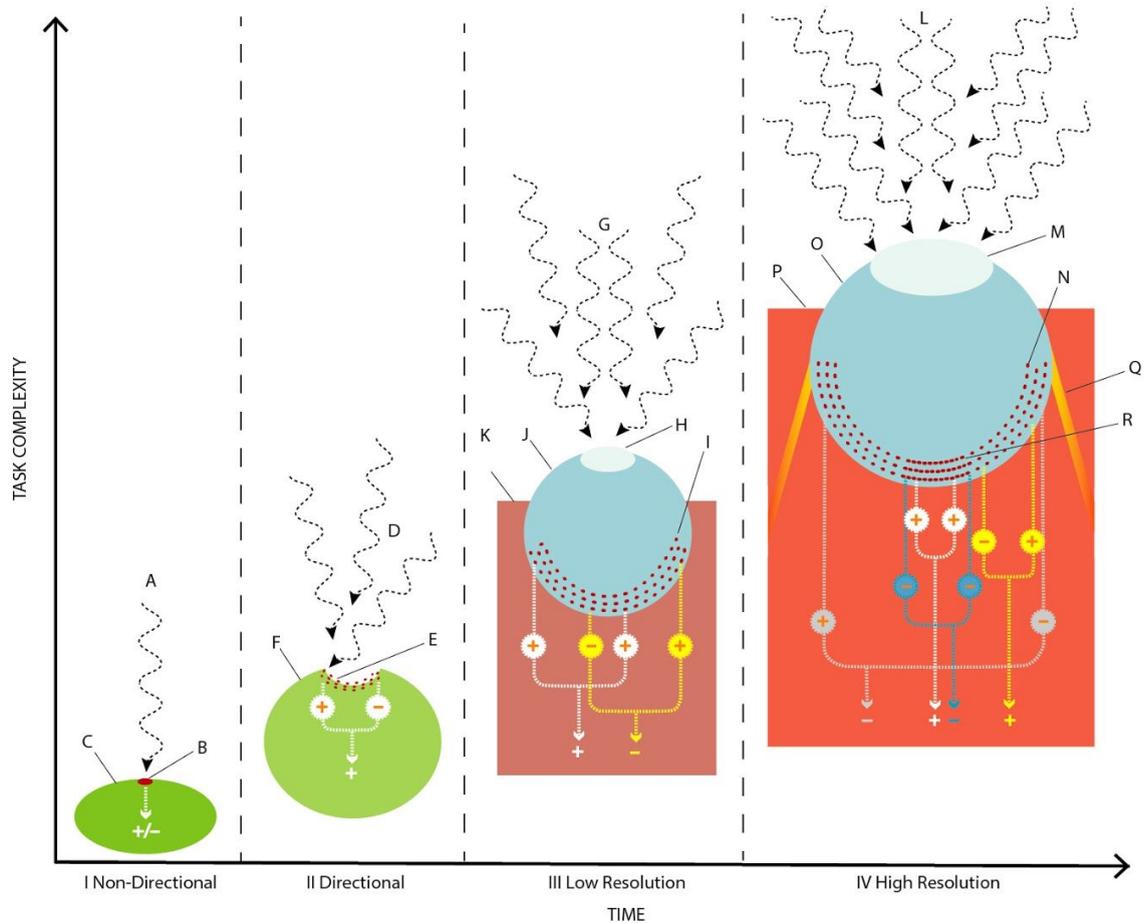

**Fig. 3.** A schematic diagram of the evolution of vision, adapted from Nilsson (2013), illustrating the three key evolutionary tendencies described above organised according to the four steps identified by Nilsson (see text), each referring to a class of visually-guided tasks. As evolution proceeds along the time axis the complexity of visually-driven tasks increases. In the I Non-Directional column, a simple organism (C) can respond only to the presence or absence of light (A), which it can capture with a photodetector (B) in only very limited amounts. In the II Directional column, greater quantities of light (D) are captured by the more complex organism (F) using an array of photoreceptors (E) that differentially respond to the direction of the light relative to the organism. These differences can be integrated within the organism by neural structures to produce a summed response that can be passed on for further processing, most likely being used to guide motion. In the III Low-Resolution column, a more complex organism (K) is able to capture still larger quantities of light (G) using a camera-type eyeball (J) and a lens (H) which focuses light onto a larger array of photoreceptors (I). The differential responses of the photoreceptors can be integrated by neural summation mechanisms and passed on for further processing. In the IV High-Resolution column, a creature of human-like complexity (P) is able to capture larger quantities of light (L) with the eyeball (O) and the larger lens (M) in order to direct the light to the photoreceptive layers (N) and in particular the densely populated array of photoreceptors at (R). Sophisticated musculature (Q) allows for rapid and accurate alignment of the eye to maximise capture of relevant light streams. The neural mechanisms responsible for integrating the incoming energy flows are commensurately more complex, as is the repertoire of behaviour this form of vision supports. The collective effect of these evolutionary tendencies has been to increase the amount of energetic differences, and differences within those differences, being sensed and processed by the visual systems of some species as their repertoire of visually-guided tasks have grown. Here the processes of differentiation and integration are shown only in the spatial dimension. In a real visual system, the temporal dimension would be equally important.

The collective effect of this evolutionary trend has been to increase the amount of energy differences, and the differences within those differences, being sensed and processed by visual systems as the repertoires of visually-guided tasks have grown.[9] This is schematically illustrated in Fig. 3.

The four steps identified by Nilsson are labelled I-IV and shown in sections on the horizontal axis in their evolutionary sequence. Each step is associated with a class of visually-guided tasks which increase in complexity over time, as indicated by the vertical axis.

In the section *I Non-Directional*, the dashed line (A) indicates light energy captured by an eyespot (B), which is part of a single-celled organism (C). (B) is capable of capturing only a small area of available light and is unable to determine its direction. The light energy is transduced at (B) and output for further processing, as indicated by the white dashed line. In this arrangement, typical of a small and relatively simple organism, light is either detected or not detected, as indicated by the +/- options. The detectable difference in the energy differences of light in this case is merely presence or absence.

In the section *II Directional*, the dashed lines at (D) indicate that the light energy captured at the photoreceptive layer at (E), located in an eye pit, is of greater quantity than in the first case and its direction can be detected by the organism (F). The photoreceptive layer is stacked, and the provision of multiple photoreceptors and processing pathways (shown in white dashed lines) allows the organism to compare responses across the photoreceptive layer (indicated by + and – symbols in discs) to determine light direction, and to integrate them into a summed response, indicated in this case by the + symbol below the arrowhead. An organism endowed with this photodetection apparatus would be able to discriminate differences in the direction of light and react accordingly.

---

[9] It is worth noting that human vision is by no means the pinnacle of eye evolution; we are visually outperformed in many ways by many other creatures, some of them with simpler visual systems (Bok, Land and Nilsson, 2012).

In the section *III Low resolution,* (G) indicates the greater quantity of light captured by the camera-type eye (J), which is focused by a lens (H) onto a stacked photoreceptive layer (I). The resulting image is projected onto a larger mosaic of receptors than in the previous case. Features of the image can be compared by processes located in the organism (K) whereby local differences at the photoreceptors can be integrated into summed responses, the number and complexity growing with the increase in numbers of photoreceptors. The output of the summing process will depend on neural circuitry involved in each case (schematically shown here). An organism endowed with low-resolution vision would have a greater ability to discriminate differences in the structure of the light commensurate with the increased complexity of its visually-guided task repertoire.

In the section *IV High Resolution*, greater amounts of light energy at (L) are captured by the larger eye (O) and focused by an improved lens (M) onto the photoreceptive layer (N), having a larger mosaic of receptors, to create a brighter and sharper image with a higher signal-to-noise ratio and higher capacity for fine discrimination. Sophisticated musculature (Q) enables the eye to line up viewed objects with the portion of the retina containing the highest concentration of receptors, the macula (R). Processes in the body of the organism (P) are able to compare the outputs from a large number of receptors and integrate them into more complex summed responses that are forwarded for further processing.[10] The design and operation of this eye is able to support the large repertoire of visually-guided tasks typical of humans, which depend on a capacity for being able to differentiate many dimensions of difference in the global light array.

---

[10] Again, this activity would conventionally be referred to as 'information processing', a description I am deliberately avoiding for reasons that are set out in Section 8.

This analysis shows that as vision evolved to guide behavioural tasks (Nilsson & Bok, 2017) it did so by exploiting the flow of light energy in the environment and the differences within that energy flow to drive useful behaviour. This resulted in the great diversity of visually-driven behaviours we witness in the animal kingdom, including those of highly socialised humans. The complexities of our motor, cognitive and social behaviours depend in large part on the capacity of our visual systems to capture sufficient numbers of photons under a wide range of lighting conditions, to finely differentiate the variations in light energy that arrive at the eyes, and to integrate these differences in ways that beneficially guide behaviour. The physical and biological operation of these principles will now be considered in more detail.

## 4. Light and the Eye

A typical illuminated environment contains one or more sources of light and a variety of materials that can scatter, absorb, reflect and refract the light. We can think of this illumination as a *lightfield* consisting of electromagnetic waves of differing intensity and frequency propagating through space. The intensity, frequency, and direction of propagation of the waves that make up the lightfield are determined by the source of illumination and the materials it interacts with. Light waves propagate from luminous sources and bounce from points on light reflective surfaces as divergent rays travelling in straight paths. These paths can also be diverted by refraction through translucent materials or by diffraction. The lightfield in a typical environment, therefore, is differentiated and the differences it contains reflect, or optically echo, the material structure of that environment.[11]

In order for the lightfield to affect behaviour, it must be observed by light sensitive organs from a particular standpoint within the environment. In humans, this process begins by selecting a sample of the lightfield that will pass through the apertures of the observer's pupils and converging the light rays to a projected image on the photoreceptors using lenses so that resulting proximal stimulus can be subjected to multiple layers of neural processing. The sample is selected by moving the eyeballs, head or body so as to align the pupils with a region of interest in the environment. Sampling the lightfield is a primary act of differentiation in vision, separating what can be seen at any moment from what cannot.

---

[11] The differentiated lightfield described here has similarities with what James Gibson calls the 'ambient light' (Gibson, 2013).

The work of moving the observer's eyes, head and body with respect to the environment does more than facilitate visual exploration. It maintains a constant state of differentiation between the incoming light and the observer's photoreceptors, even when fixating on a point in space. When an eye is exposed to a light stimulus that is stabilized on the retina the corresponding percept rapidly fades (Heckenmueller, 1965). Continuous rapid involuntary movements made by the eyes (microsaccades) ensure that photoreceptors are exposed to an ever-changing pattern of stimulation. This points to a basic principle of vision, which is that we primarily perceive *difference*; we tend to become oblivious to that which does not change, or which we fail to notice change (Gibson, 2013).[12]

Besides the extraocular muscles that control the saccades and the direction of gaze are various muscles that adjust the diameter of the pupil and the shape of the lens. Adjusting pupil diameter controls the quantity of light entering the eye and the amount that is directed through the centre of the crystalline lens where optical quality is highest, both of which affect the focal properties of the retinal image. Altering the shape of the lens during accommodation changes the refractive properties and the focal length of the eye. Correct focal length results in a sharply focused image on the foveal region of the retina where photoreceptors are most densely arrayed. The product of the work performed by eye muscles is a retinal image of maximal sharpness for given viewing conditions. Sharply focused retinal images enhance the ability of the observer's visual system to differentiate the structure of the lightfield with the level of acuity needed to support their behavioural tasks (Land & Nilsson, 2012).

## 5. Photoreceptors and Energy

Each human retina contains approximately 100 million photoreceptors. When light waves interact with chromophores in photoreceptors, they pass on their (higher) energy in discrete quanta (Pirenne, 1967). The energy quanta, or photons, perform work on the photoreceptors by causing retinal molecules embedded in the rhodopsin (the photosensitive pigment found in rods) to change shape. The impact of the photon momentarily knocks an electron belonging to a retinal carbon atom into a higher energy orbit, so weakening its bond with a neighbouring atom and flipping the molecule from the 11-*cis* to the all-*trans* configuration. The transduction of light energy to metabolic energy is then enacted through a cascade of chemical reactions in the photoreceptor that, when powered by chemical energy supplied to the retina through blood vessels, causes the exchange of electrochemical energy between retinal neurons that drives the work of early visual processing.

---

[12] The proximal stimulus is always moving in vision, whether or not the distal stimulus is.

The light energy that interacts with individual photoreceptors can differ in two ways, each of which causes a different impact on photoreception and subsequent visual processing:

a. in the frequency of the waves occurring in the electromagnetic field per unit of time or, equivalently, by the wavelength, where the higher the frequency or the shorter the wavelength the greater the energy, and

b. in the intensity of the waves occurring in the electromagnetic field, where the greater the intensity the greater the energy.[13]

Cones are the colour sensitive photoreceptors in the eye. There are three types, each having a different absorption spectrum that determines its sensitivity to different wavelengths or energies of light: long-wave (L), medium-wave (M) and short-wave (S). Each absorption spectrum corresponds to a different perceived range of colour. L-cones have a peak sensitivity that is tuned to the red-yellow part of the visible electromagnetic spectrum, M-cones to the yellow-green part, and S-cones to the blue-violet part.

The processing of colour in the visual system works by integrating neural responses from the different types of cone in order to compare the relative degree of stimulation between them rather than detecting absolute wavelength values on a cone-by-cone basis. This is an example of a principle found widely in visual processing in which the structure of the lightfield absorbed by photoreceptors is first differentiated into discrete neural responses and then integrated by neural processes of summation or comparison (Kandel, 2013).

---

[13] There is evidence that humans can also discriminate the polarization of light, although the contribution this makes to guiding behaviour is not clear (Temple et al., 2015).

A typical lightfield also varies in the intensity of the energy it carries. The level of light intensity affects the rate at which photoreceptors are stimulated; more intense light will stimulate more photoreceptors more frequently, leading to faster rates of neural firing in retinal ganglion cells (Milosavljevic et al., 2018). This in turn affects the perceived brightness of the resulting percept. A typical percept will be composed of regions of varying brightness that are causally related to the structure of the lightfield, and hence of the environment.[14]

Like the cones, rods are not independent light detectors. Communication via electrical couplings between rods, and between rods and cones, allows for integration of responses of dispersed photoreceptive cells, meaning that the work of visual processing—biologically differentiating and integrating differences in the lightfield—is beginning even as the light energy is being transduced (Copenhagen & Owen, 1976).

## 6. From Eye to Brain

The retinal ganglion cells that project from the eye to the brain are very small in number compared to the population of photoreceptors they serve. This means that a large amount of processing of the light stimulus must be undertaken among post-photoreceptor cells—bipolar, amacrine, and horizontal—as well as in ganglion cells so that the most behaviourally useful patterns of neural activation are streamed through the optic nerve to the visual processing areas in the brain.

The early visual system is especially sensitive to differences or contrasts in brightness across the visual field [15] and will actively work to enhance these differences in order to increase the probability that they will be perceived. This is evident in the organization of retinal ganglion cells containing centre and surround receptive fields that are sensitive to ON/OFF differences in illumination within a local region rather than to the global presence or absence of light. In the case of an ON cell, for example, excitation will increase if light hits the centre of the receptive field and decrease when it hits the surround. The response is also weak when light hits both the centre and the surround. For OFF cells the inverse applies. This antagonistic organization means that contrasts of intensity within the portion of the light field projected to the retina will tend to stimulate neighbouring cells while areas of homogenous illumination will not (Kuffler, 1953). Some ganglion cells respond selectively to differences in the direction and speed of stimuli across their receptive field caused by changes in the structure of the lightfield as objects in the environment move (Barlow et al., 1964).

---

[14] Note that while the structure of a visual percept, the light field structure, and the environmental structure are causally related, they are not identical. There are many cases in visual perception where phenomena are perceived that are not present in the retinal image or the environment, and vice versa (Gregory, 2015).
[15] The visual field is the total area of the visible world available to both eyes when fixating on a given point in space. Gibson describes it as the "momentary patchwork of visual sensations" (Gibson, 2013).

Enhancement of contrast between differences in the structure of the lightfield is further facilitated by lateral inhibition processes operative among horizontal and amacrine cells in the retinal layer. The excitatory influence of light on a photoreceptor can be counteracted by the inhibitory influence of other photoreceptors, connected via horizontally-aligned cells, which has the effect of suppressing activity in remote receptive fields so that local fields can generate a stronger relative response. These are examples of visual processes that work by integrating differences through a vertical-horizontal organization of cellular interactions in order to accentuate states of difference between retinal cells. The same fundamental principles can be found at work at higher levels of visual processing (Kandel, 2013).

## 7. Energy Flow in Neurons

Visual processing is carried out mainly by neurons organized into networks and functionally specific structures within the brain that affect each other electrochemically.[16] Neural behaviour can be described in terms of energy flow, which is expedited by the inter- and intra-neural exchange of kinetic energy (motion) and the performance of work (displacement by application of force).[17] Individual neurons import differences of electrochemical energy from adjacent cells and integrate those differences within the cell body in such a way that they either export differences to other adjacent cells—for example by firing an action potential—or do not. The importation of energy differences into a retinal ganglion cell, for example, occurs at sites on its dendrites, where synapses from adjacent amacrine and bipolar cells terminate. The wave of polarization in the excited presynaptic cell moves towards the synapse, forcing voltage gated calcium channels to open and initiating an influx of positively charged ions across the chemical gradient. This in turn causes neurotransmitter bearing vesicles in the presynaptic neuron to move towards the synaptic cleft and fuse with the local membrane.

---

[16] Besides neurons, of which there are several kinds, the nervous system contains various glial cells, such as astrocytes, that contribute to neurobiological processes in ways that are still not fully understood.
[17] This is in contrast to the widespread tendency to describe neuronal functioning in terms of information exchange or signal transmission.

The vesicles open to release neurotransmitters into the synaptic cleft, and these migrate towards the postsynaptic membrane where they bind to the membrane surface and trigger the opening of local ion channels. These channels allow ions to flow in, which alters the voltage or potential within the postsynaptic neuron. If the effect of neurotransmitter release is to allow negatively charged ions to flow into the postsynaptic cell, then it will be further polarized and so more likely to inhibit a response, preventing exportation of energy differences. If neurotransmitter release allows positively charged ions to flow in, then it will be depolarized and so more likely to excite a response in the postsynaptic neuron and export energy differences to other adjacent cells. Even from this elementary-level account it is evident that neurons perform work by regulating the flow of kinetic events or *movement*, from the propagation of the action potential, the opening and closing of the voltage gated channels, the relocation of the vesicles, the migration of neurotransmitters, to the influx of ions into the postsynaptic cell.

All this neuronal motion is driven by electrical potentials and chemical gradients, both within the cell and between the cell and its surroundings. These energetic disequilibria create a physical tension (in the form of a voltage or polarity) that can be exploited to do the work of moving matter by force, examples of which have been mentioned. In their resting phase, neurons have a polarity of about 65-70mv across the interior and exterior of the semi-permeable cell membrane due to the difference in concentration of ions suspended in fluids. The cell must do continual (and energetically expensive) work to counter the natural tendency of the gradient towards entropic equilibrium (Harris et al., 2012), including employing nano-scale pumps to move ions against gradients.

The energy required to recharge these gradients once they are levelled, and to drive other movement in the cell, is derived from molecules of adenosine triphosphate (ATP) that are mainly synthesized in mitochondria (Harris et al., 2012; Sperlágh & Vizi, 1996).[18] During the action potential phase, an energetic difference across the neuron membrane is propagated from the trigger zone in the cell body, along the axon to the synapses that terminate the axon. [19] This action takes the form of a wave of alternating depolarization and repolarization which culminates in an alteration of the ion balance in the fluids in the axon terminals and movement of ion gates, pumps and neurotransmitter vesicles described above. In addition to regulating and propagating these spatial differences, neurons also generate temporal differences due to variations in the timing and frequency of action potentials (Dan & Poo, 2004; Tiesinga et al., 2008). These differ in response to imported energy differences and, in turn, differentially affect neighbouring cells through exported differences. In sum, neurons expedite energy flow by propagating and regulating differences of energy, both spatial and temporal, through the cell body, and from cell to cell, exploiting the motive power inherent in electrochemical energy differences.

## 8. Energy Processing in Neurons

The work of integrating, or summing, the many energetic differences received by a neuron from its presynaptic connections into a single neural response can be regarded as a form of *energy processing*, as schematically illustrated in Fig. 4. Panel A(1) represents a ganglion cell body with a semipermeable membrane containing positively and negatively charged ions, and A(2) represents multiple presynaptic connections from retinal cells. The extracellular fluid in A(3) also contains positively and negatively charged ions, and the voltage potential across the cell membrane is such that the cell is at rest and the trigger zone (shown in blue) at A(4) is not activated.

---

[18] ATP contains energy due to the relative instability of the chemical bonds between the phosphates, specifically their high concentration of negatively-charged electrons that tend to repel one another, resulting in a tension or energy difference within the system, much like the water pressing against the dam in Fig. 1.2. When in contact with a suitable enzyme sufficient additional energy is acquired to break one of the phosphate bonds in the ATP, allowing the energy difference in the system as a whole to equalize, while the energy that is released (or converted) through this process can be harnessed to perform useful biological work elsewhere. To extend the analogy with the dam: if a certain amount of energy is used to breach the dam wall a much larger amount of energy is released (converted from potential to kinetic) as the water levels out.

[19] The action potential is well named, capturing both the active, kinetic aspect and the differential, tensional aspect of this neural behaviour. According to the Oxford English Dictionary, the term first occurs in English in 1913, having been employed by an Italian physiologist, Galeotti, in 1904 to distinguish it from the 'observed' potential of the cell at rest. The impulsive nature of the action potential is also implicitly recognised by the commonly-used terms 'firing' and 'spiking', the latter referring to the abrupt peak registered by an encephalographic graph in response to an energetic difference in neural activity.

In panel B, energy differences move towards the cell from presynaptic connections, in this case more positively charged than negative. As shown in panel C, this causes an increase in the relative number of positive ions in the cell, which alters the voltage potential across the cell membrane. Once this reaches a certain threshold, as shown in panel D, the trigger zone at D(1) is activated (shown as pink) causing an energetic difference to be exported from the ganglion cell (pink arrow) as an action potential.

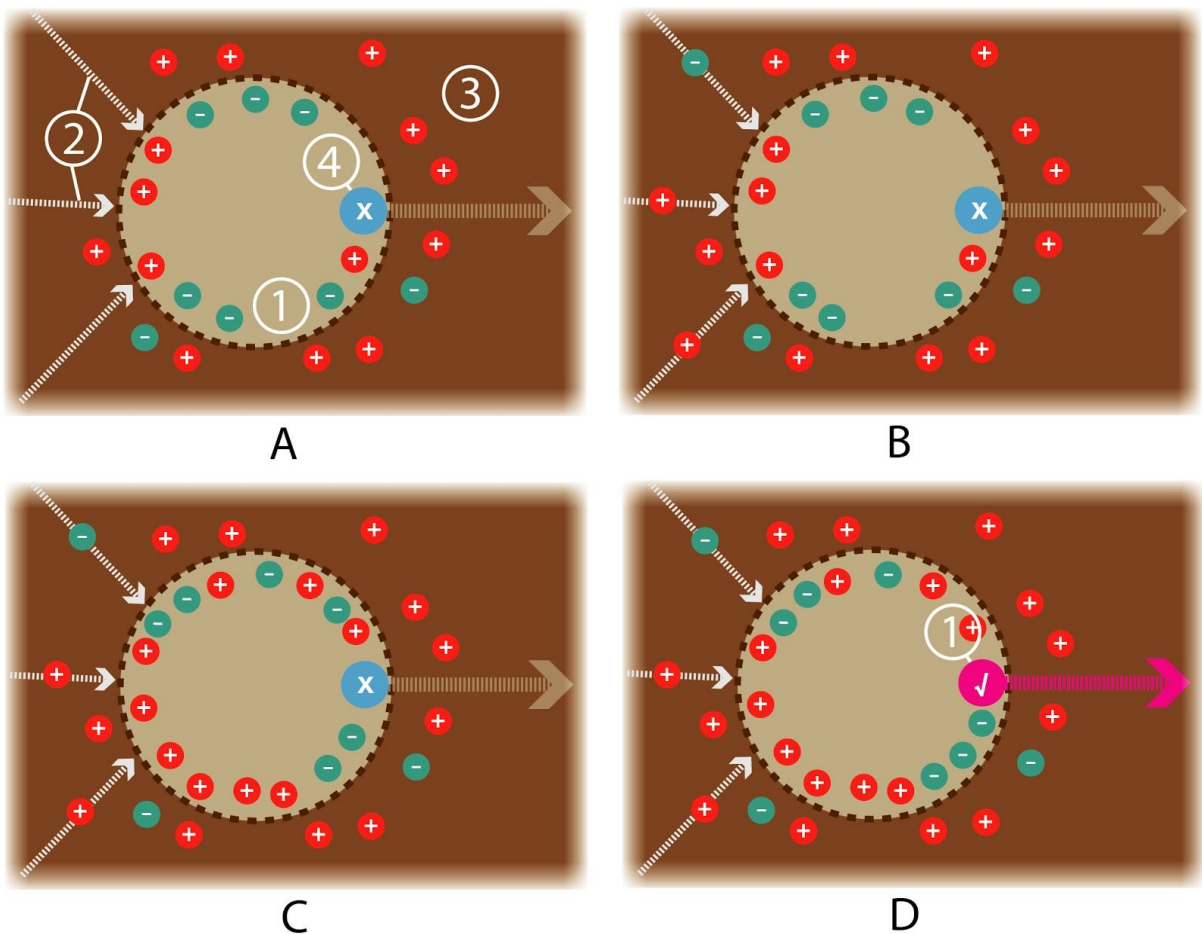

**Fig. 4.** A schematic diagram of energy processing (summation) in a retinal ganglion cell. In Panel A, the neuron (1) contains ions that differ in polarity from those in the extracellular fluid (3) so creating a potential, or energy gradient, across the cell membrane. The cell is in contact with retinal cells via dendrites (2) and a trigger zone (4) is inactive, meaning the cell is at rest. In Panel B, energy differences, in the form of positive and negative charges, are imported from retinal cells via the dendrites. In Panel C, the relative concentration of charge across the cell membrane is altered by the importation of energy differences, thus altering the potential across the membrane. In Panel D, the change in potential is sufficient to activate the trigger zone (1) resulting in the exportation of an action potential (in pink). In this simplified description, we can see how the cell dynamically integrates various energetic differences into a summed response. In real neurons, summation occurs both temporally and spatially.

In information-driven accounts, this activity might be described in terms of 'information processing' or 'neural computation' and the action potential might be represented as '1' in binary code while the non-active state of the neuron would be represented with '0' (Sterling & Laughlin, 2015). Treating neurons as processors of information is useful if one wishes to mathematically or computationally model the organisation or behaviour of the energetic differences in a system, such as the difference between a neuron firing and not firing. And information theoretical tools can be powerfully applied to the analysis and modelling of the complex differences of differences, or integrated differences, that make up the structures and behaviours of biological systems (Delgado-Bonal & Martín-Torres, 2016; Koch et al., 2006; Milosavljevic et al., 2018; Tiesinga et al., 2008).

However, it is doubtful whether treating neurons as information processing units is an accurate characterisation of their biological purpose. As previously noted, the term 'information' is variously and not always consistently applied in science, having at least three main uses: the *colloquial*, referring to news, knowledge, or meaning that is acquired or transmitted (Bates, 2009; Bateson, 2000), the *mathematical*, referring to the amount of knowledge that a measurer has about the predictability of a system being measured (Adami, 2016; Shannon, 1948), and the *computational*, in which differing states of a physical system are assigned (usually binary) numerical values for the purposes of computation ((Sterling & Laughlin, 2015).[20]

---

[20] Each of these definitions is broadly exemplified in the quotes about visual information cited in the first paragraph of the introduction: Marr and Palmer use information in the first sense, Koch et al. use it in the second sense, and Sterling and Laughlin use it in the third sense. It is worth noting that the physical states represented by computational information are usually differences of energy, as in differences of voltage in the transistors in computer chips, or pulses of light in optical communications devices.

If we consider the biochemical activity of a single neuron, it is hard to see how any of these definitions apply literally. To say that a neuron is transmitting information within itself or to other neurons is either to say that it is conveying news, knowledge, or meaning—which can only be true in a metaphorical sense since an individual neuron is unlikely to have the capacity to entertain semantic content—or is passing on a value derived from a measurement of how uncertain the measurer is about the predictability of a system—which cannot be so since neurons are not measuring agents—or is generating and communicating binary numerical values—which is inaccurate since individual neurons neither generate nor transmit numbers. Information in any of these commonly used senses is not, therefore, a property that individual neurons possess or can process, other than metaphorically. This limits the fundamental explanatory power of approaches based on treating neurons as information processors.[21]

In this energy-driven account, neurons are primarily dynamic processors of energy differences. Neural activity constitutes a form of biological work that is driven by electrochemical gradients in which various kinetic and tensional differences (in the form of mobilized ions) are imported from multiple sources, integrated through summation of differing electrical charges, and then exported, when appropriate conditions are met, through the action potential.[22]

## 9. Further Visual Processing

Having processed and integrated the energetic differences imported from their respective populations of retinal cells, many of the ganglion neurons transport their action potentials along the optic tract to the lateral geniculate nucleus (LGN) in the thalamus. Just as the ganglion cells act to integrate the various energy differences they import from retinal layer cells using a centre-surround organization, so the LGN integrates the differences it imports from many ganglion cells, again using a centre-surround organization. In particular, cells in the LGN are highly sensitive to patterns of difference containing edges or line contrasts, and such features will tend to be accentuated in the patterns of energetic differences that are conveyed from the LGN to the visual cortex.[23]

---

[21] Vast assemblies of neurons, working in concert, can acquire, process, and transmit information in all three senses since the sentient cogitation of information is a common part of mental life, and one that is presumably generated by neural activity. But precisely how the mass activity of individual neurons produces subjectively experienced meaning, knowledge, acts of measurement, or numerical values remains a mystery, and is beyond the scope of this paper.

[22] It is sometimes inferred from the all-or-nothing nature of the action potential that the neuron is a digital information relay or computational unit that signals a binary code of ones and zeros. Many neurons in the retina share graded potentials, which are more analogue than digital, and in all neurons both receptor and synaptic potentials are graded, as are the integrative processes occurring in the cell that determine whether a firing event is triggered. There is still much to learn about the complex behaviour of neurons; see recent discoveries about graded calcium-mediated dendritic action potentials (Gidon et al., 2020), and even in the early 1950s researchers were cautioning about inferring a 1:1 correspondence between cell discharges patterns and information (Kuffler, 1953).

The combined effects of differentiation and integration occurring at this level of visual processing, as in many others, is to produce patterns of neural activity that represent relatively primitive differential elements, such as lines, edges, movements, and contrasts of light and dark that will ultimately be perceived by the observer as parts of objects. The precise brain regions and neural mechanisms that support the conscious visual perception of objects have not been fully identified, but activity in the temporal and prefrontal cortical areas is generally thought to be of central importance (Panagiotaropoulos et al., 2014).

## 10. Energy Flow in Vision and Behaviour

This brief sketch of the evolution of vision, the behaviour of light, photoreception in the eye, and the early stages of visual processing highlights some of the fundamental principles through which vision works as a physical, chemical and biological process. According to this account, visual systems can be regarded as dynamic processors of energy that exploit differentiated properties of the lightfield in order to actuate and guide behaviour.

Energy, forces and work manifest differences of motion and tension, as discussed in Section 2, while light is also differentiated in terms of wavelength, intensity and direction of propagation. Light energy acts upon photoreceptors by changing the configuration of photosensitive compounds, thus allowing photoreceptors to register differences in the received portion of the lightfield. This triggers streams of electrochemical activity in the eyes and brain in which vast arrays of action potentials propagate from ganglion cells to intermediate and higher level visual processing regions. These streams of energetic activity are further differentiated and integrated into patterns of neural excitation and inhibition, some of which are relayed to motor control regions of the brain and, ultimately, drive a behavioural response to the visual stimulus.[24]

---

[23] Substantially more neurons project from the visual cortex to the lateral geniculate nuclei than from LGN to visual cortex; visual processing is not unidirectional but involves rich feedback connections among many brain regions.

Behaviour is itself a form of energy-driven work carried out by moving parts of the body through the actuation of antagonistically arranged muscles. In fact, it is not too simplistic to think of the entire sensory, nervous and muscular apparatus as a fantastically complex integrated system for converting differences of environmental energy (light being only one kind) into differences of metabolic energy in the body and then into differences of kinetic energy in the form of behaviour, constituting an energy flow as schematically illustrated in Fig. 5.

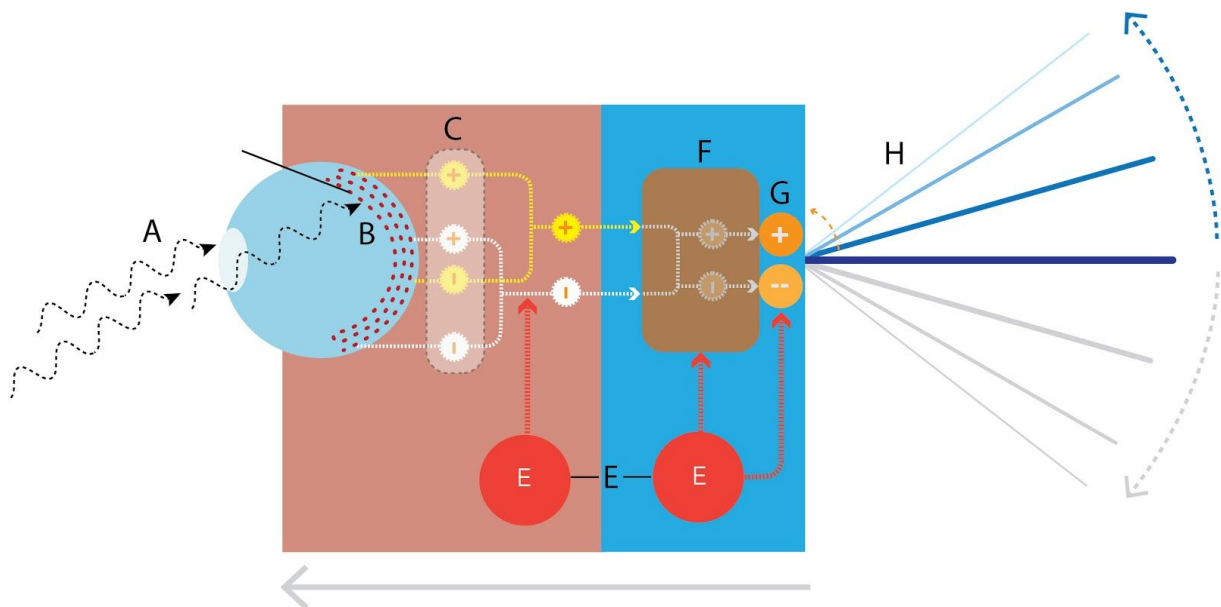

**Fig. 5.** A schematic diagram of energy flow in vision and behaviour. This idealized creature absorbs varying patterns of kinetic energy from light through the eye (A), which does work on its photoreceptors (B) so transducing the energy into other kinetic and potential forms of energy. Work is then done by biochemical processors (C) that compare the differentials across the array of photoreceptors, using energy potentials and gradients to perform this work, and integrate them into summed responses (D) that reflect the intensity and direction of the light. Using energy supplied from the organism's own resources (E) the integrated differences are exported to a motor control region (F) where they are further integrated and differentially excite effector cells (G) that trigger the controlled movement of a flagellum (H) that works by applying force to move the creature, so converting the

---

[24] I am omitting for the sake of space any discussion of 'top-down' effects, feedback pathways, recurrent activity, or predictive coding mechanisms in vision that allow higher level and later visual processes in the brain to influence lower level or early visual processes.

metabolic energy into kinetic form. The behaviour of the creature as a whole is an example of biological work that is powered by external and internal energy flows, the overall purpose of which is to sustain its existence.

The propulsive nature of energy and forces as they actuate work in biological systems—their capacity to drive motion and behaviour—stands in contrast to information in the following way: information, taken in its most rigorously defined Shannonian sense as mathematical information, has no intrinsic causal power (Shannon, 1948). It cannot dynamically move or change things in the physical world since it is a numerical value derived from a process of measurement undertaken by a suitably skilled measurer; it is a piece of knowledge *about* a system rather than a property of a system *itself*.[25] This limits the scope of information-based approaches to fully explain the dynamical causes of vision and visually-driven behaviour.

## 11. Conclusion

This paper has drawn attention to the way energy flow drives the fundamental physical, chemical and biological processes of early vision. Although a great deal more work is required to turn this initial sketch into a detailed picture of the larger visual system, I suggest that even the elementary analysis presented here can contribute to our understanding of the fundamental nature and purpose of vision:

1. It specifies general principles operative at a fundamental level of vision,

2. It helps to explain the role of energy flow in the evolution and operation of vision,

3. It avoids some of the limitations of the information-driven approach.

Looking forward, perhaps one of the most potentially important aspects of this energy-driven approach is the contribution it could make to our understanding of how certain physical, chemical and biological dynamics in the eyes and brain can lead to our having a conscious experience of seeing the world. Elsewhere I have proposed that energy, along with forces and work, not only plays a critical role in driving the metabolic processes on which neural activity depends but also in generating phenomenal experience itself (Pepperell, 2018). If we find that the principles identified in the present analysis of early visual processing can be extended to the higher-level neural processes involved in the phenomenology of visual perception, cognition, and consciousness—and assuming such phenomenology is not generated by some other kind of neural activity not yet known to us—then these features of our mental life would also be explicable as the product of the dynamics of energy, forces and work.

---

[25] To clarify this point with an analogy: I may measure the temperature of the sun's core to be 15 million degrees Celsius, but my knowledge of this value belongs to me and not to the sun. To extend the analogy: 15 million degrees Celsius may be a large value, but it is not hot.

## Acknowledgements

## Funding

This research did not receive any specific grant from funding agencies in the public, commercial, or not-for-profit sectors.

## Conflicts of Interest

There are no competing or conflicting interests of relevance to this article.

## Ethics

Not applicable.